\documentclass[aps,prl,twocolumn,superscriptaddress,showpacs,preprintnumbers]{revtex4-2}

\usepackage{amsmath,amssymb,amsfonts,bm}
\usepackage{graphicx}
\usepackage[colorlinks=true,citecolor=blue,linkcolor=blue,urlcolor=blue]{hyperref}

\newcommand{\PF}{\mathrm{PF}}

\newcommand{\epsb}{\varepsilon}
\newcommand{\Tau}{\tau}

\begin{document}

\title{Exact Solvability via the KP Hierarchy for $\beta=L^2$ Random Matrix Ensembles}

\author{Christopher D.\ Sinclair}
\affiliation{Department of Mathematics, University of Oregon, Eugene, OR 97403, USA}

\date{\today}


\begin{abstract}
Random matrix ensembles with Dyson index $\beta=L^{2}$ describe systems of $M$
charge-$L$ particles interacting logarithmically in the presence of an external
potential, yet exact formulas for their physical observables have remained
elusive for $L\neq 1,2$. We show that, for $L$ even, $\beta=L^{2}$ ensembles are
governed by the KP hierarchy at finite particle number—paralleling the KP
solvability of classical $\beta=1,2,4$ ensembles. The partition function is a
hyperpfaffian $\tau$-function satisfying the Hirota bilinear identity, and
correlation functions are generated by finite-order differential operators
acting on this $\tau$-function. The key mechanism is an emergent quantized
momentum that stratifies the system into discrete sectors, enforcing momentum
conservation as a selection rule. This produces a dramatic dimensional reduction
from ${LM\choose L}$ to $O(L^{2}M)$, enabling explicit computation of physical
observables. 
\end{abstract}

\maketitle
\textbf{Introduction.}
Random matrix theory has become a foundational tool in mathematical physics,
with deep impact on quantum chaos, disordered systems, numerical analysis, and
number theory \cite{Dyson1962,Mehta2004}.  Classical results show that many
distinct physical systems share the same local spectral statistics, a phenomenon
known as universality \cite{Forrester2010}.  The Dyson ensembles with
$\beta=1,2,4$ play a special role: their determinantal or Pfaffian structure
allows exact evaluation of correlation functions and connects them to integrable
hierarchies, orthogonal polynomials, and representation theory
\cite{DJKM1982,JimboMiwa1983,MiwaJimboDate2000,HarnadOrlov2021}.  The appearance
of integrable hierarchies in random matrix theory reflects a broader algebraic
phenomenon, long recognized in the theory of matrix integrals and Toda/KP
lattices \cite{AdlerVanMoerbeke2001}. These ensembles serve as a benchmark for
what a ``solvable'' random matrix model looks like. 

Away from the classical values $\beta=1,2,4$, very little algebraic structure
survives.  For general $\beta$, the joint density retains the log-gas form
\[
F(x_1,\dots,x_M)\propto \prod_{i<j}|x_i-x_j|^\beta
   \prod_{k} e^{-V(x_k)},
\]
but no determinantal or Pfaffian representation exists, correlation functions
lack simple closed descriptions, and available methods---such as tridiagonal
models \cite{DumitriuEdelman2002}---are perturbative or asymptotic
\cite{RamirezRider2011}. A central question has been whether there exist
nonclassical values of $\beta$ for which exact algebraic structure is restored.

In this Letter, we show that $\beta=L^{2}$ ensembles with $L$ even are exactly
solvable via the KP hierarchy. The partition function is a hyperpfaffian
$\tau$-function satisfying the Hirota bilinear identity, with all correlation
functions generated by differential operators acting on this $\tau$-function.
The solvability arises from an emergent quantized momentum that enforces a
conservation law. Momentum conservation acts as a selection rule, collapsing the
effective dimension from $\binom{LM}{L}$ to $O(L^{2}M)$ and allowing for exact
formulas for physical observables. 

Algebraically, the interaction term $|x_i-x_j|^{L^{2}}$ is identified exactly
with the exterior volume spanned by $L$-fold Wronskian blades, lifting the
ensemble into an exterior algebraic framework where hyperpfaffians (polynomial
invariants of $L$-forms) replace determinants. The partition function becomes
the hyperpfaffian of a Gram $L$-form, the analog of antisymmetric Gram matrices
that appear when $\beta=1,2$, and time deformations of this hyperpfaffian define
a $\tau$-function.  

Time derivatives act by shifting momentum amplitudes, and pairing particle
insertion and annihilation subject to momentum-conserving Pl\"ucker relations
yields the Hirota bilinear identity. This identity encodes the integrable
hierarchy satisfied by the $\tau$-function.

The combination of hyperpfaffian geometry, momentum conservation, and integrable
structure establishes $\beta=L^{2}$ ensembles as a new solvable class of
nonclassical random matrix models, extending exact solvability beyond the
classical Dyson indices ($\beta=1,2,4$). They occupy a middle ground between the
fully solvable classical ensembles and the analytically intractable generic
log-gases, demonstrating that integrability can emerge in nonclassical regimes
through algebraic mechanisms. The framework reveals that exact solvability is
not limited to the classical Dyson indices, but extends to an infinite family of
square-$\beta$ Dyson indices. We focus on $L$ even as it simplifies the
presentation; the case of odd $L$, which involves a related formalism based on
$2L$-forms, will be addressed in a future publication.

\textbf{Main Result.} 
For even $L$ and $\beta = L^{2}$, the $M$-particle $\beta$-ensemble admits a
canonical representation in a finite-dimensional exterior algebra, in which the
partition function is a hyperpfaffian $\Tau(t)$ depending polynomially on
deformation parameters $t=\{t_k\}$.  This hyperpfaffian defines a KP
$\tau$-function: it satisfies the Hirota bilinear identity
(with Miwa shifts weighted by $L^{2}$), and all correlation functions are
generated by Miwa shifts or by differential operators acting on $\Tau$.
The integrable hierarchy emerges directly from Pl\"ucker relations on a
momentum-graded subspace determined by the interaction $|x_i-x_j|^{L^{2}}$,
establishing $\beta=L^{2}$ ensembles as a solvable nonclassical class
of random matrix models. All integrable identities derived here arise
from finite-dimensional Pl\"ucker geometry at fixed particle number, without
invoking infinite-rank Grassmannians or large-M limits.


\textbf{Wronskian geometry.}
Particle locations are identified with the Grassmannian of $L$-dimensional
subspaces of an $LM$-dimensional vector space $V$, encoding the $L$-body interaction
as an antisymmetric geometric object. For each point $x$, define the
$L$-blade (decomposable element)
\[
\omega(x) = v(x)\wedge v'(x)\wedge\cdots\wedge v^{(L-1)}(x)
\in \Lambda^{L}V,
\]
where $v$ is the vector of monomials generating the space of polynomials of degree
less than $LM$, and primes denote derivatives.  The coordinates of $\omega(x)$
are indexed by $L$ ordered integer quantum numbers $(u_1,\ldots,u_L)$, with
amplitude given by the Wronskian determinant
$\mathrm{Wr}(x^{u_1},\ldots,x^{u_L})$. This Wronskian is a monomial with power
$u_1 + \cdots + u_L - L(L-1)/2.$ Recentering these powers by the canonical sum
$\overline{\Sigma} = L(LM-1)/2$ defines the momentum $p_u = u_1 + \cdots + u_L -
\overline{\Sigma},$ which grades the coordinates of $\omega(x)$ into
momentum sectors. 

The key algebraic identity is the confluent Vandermonde formula
\[
\star(\omega(x_1)\wedge\cdots\wedge\omega(x_M))
= \prod_{i<j}(x_j-x_i)^{L^{2}},
\]
where $\star$ denotes the Hodge dual (projection onto the determinantal line).
This identity shows that the entire $L^{2}$-interaction structure arises 
from an antisymmetric multilinear object.

\textbf{Hyperpfaffian representation.}
Integrating the coordinates of a Wronskian blade against the
probability measure $\mu$ produces the Gram $L$-form
\[
\gamma \;=\; \int \omega(x)\, d\mu(x),
\]
the direct $L$-form analogue of the antisymmetric Gram matrices in 
$\beta=1,4$ ensembles. Using the confluent Vandermonde identity, the
$M$-particle partition function is
\[
Z \;=\; \star\!\left(\frac{\gamma^{\wedge M}}{M!}\right)
\;=:\; \PF(\gamma),
\]
the hyperpfaffian of $\gamma$~\cite{LuqueThibon2002,Sinclair2012}, encoding the entire
log-gas interaction in the antisymmetric structure of $\gamma$.

Momentum conservation emerges from the geometric structure. Each blade 
$\omega(x)$ decomposes into components of definite momentum $p$, and wedge 
products add momentum labels. The determinantal line carries total momentum 
zero, and the Hodge projection $\star$ selects only the momentum-zero component 
of $\gamma^{\wedge M}$. Consequently, \emph{only configurations with net 
momentum zero contribute to $Z$}. This momentum-conservation law is the 
mechanism for dimensional reduction: the hyperpfaffian acts as a momentum 
filter, collapsing the $\binom{LM}{L}$-dimensional exterior algebra onto an 
$O(L^{2}M)$-dimensional subspace. All physical observables—partition 
functions, correlation functions, and time flows—respect this conservation 
law.

\textbf{Momentum spine.}
Momentum conservation organizes the exterior algebra into finitely many discrete
sectors, and we can define a set of fundamental $L$-forms, $\{\epsb_p\}$ indexed
by momentum. The linear span of $\{\epsb_p\}$ is the {\em momentum spine}. In
this basis,
\[
\omega(x) = \sum_{p} x^{P+p}\,\epsb_p, \qquad \gamma = \sum_p m_p \epsb_p,
\]
where here $P = L^2(M-1)/2$ is a canonical power which compensates for the
Wronskian derivative shift, and $m_p = \int x^{P+p}\,d\mu(x)$ are the shifted
moments. Allowed momenta are $|p| \le L^{2}(M-1)/2$, giving effective dimension $O(L^{2}M).$

\textbf{The hyperpfaffian $\tau$-function.}
To connect with integrable hierarchies, we introduce time deformations of the
measure, extending the static partition function $Z$ to a time-dependent
$\tau$-function. Given times $t=\{t_k : k\in\mathbb{Z}\}$, introduce the
time-deformed measure
\[
d\mu_t(x)=e^{\xi(x,t)}\,d\mu(x), \qquad \xi(x,t) = \sum_k t_k x^k.
\]
The moments become time-dependent, yielding the time-deformed Gram $L$-form
$\gamma(t) = \sum_p m_p(t)\,\epsb_p.$ The time-deformed partition function
\[
\Tau(t)=\PF(\gamma(t))
\]
is a finite-dimensional hyperpfaffian serving as the $\tau$-function of the
$\beta=L^2$ ensemble. From the probabilistic point of view, the $\tau$-function
packages all correlation functions into a single generating function. It plays
the same structural role as KP $\tau$-functions
\cite{Sato1981,JimboMiwa1983,MiwaJimboDate2000}, with hyperpfaffians replacing
determinants or ordinary Pfaffians. 

\textbf{Time derivatives and momentum flows.}
Time derivatives shift moments. Writing $\partial_k := \partial_{t_k}$,  
we have $\partial_k \gamma = \sum_p m_{p+k}(t)\,\epsb_p$, so $\partial_k$ acts 
as a shift operator on the momentum spine with composition rule 
$\partial_k\partial_\ell\gamma = \partial_{k+\ell}\gamma.$ 

The time derivatives of $\Tau$ follow from hyperpfaffian multilinearity:
\[
\partial_k \Tau
  = \star\!\left(\partial_k \gamma \wedge \frac{\gamma^{\wedge (M-1)}}{(M-1)!}\right),
\]
shifting momentum in one factor $\gamma$. 

Polynomial differential operators on Gram forms lift canonically to operators 
on the $\tau$-function. For a polynomial $Q(x)=\sum_k q_k x^k$, define
\[
Q(\partial)[\gamma] := \int Q(x)\,\omega(x)\,d\mu(x) ,
\qquad Q(\partial):=\sum_k q_k\,\partial_k,
\]
where $\partial_k$ acts under the integral as multiplication by $x^k$.
$Q(\partial)$ acts as a Toeplitz operator on the momentum spine. 

Since $\Tau=\PF(\gamma)$ is a homogeneous degree-$M$ polynomial in the moments
$m_j(t)$, the operator $Q(\partial)$ lifts via polarization to a differential
operator $\widehat Q(\partial)$ satisfying the fundamental identity $\widehat
Q(\partial)[\Tau] = \mathrm{PF}(Q(\partial)[\gamma]).$ This is the hyperpfaffian
analogue of vertex-operator calculus in KP theory.


\textbf{Correlation functions.}
Physically relevant observables of the ensemble include the $m$-point correlation functions $R_m(\mathbf{y})$, 
which represent the probability density of finding particles at locations $\mathbf{y}=(y_1,\dots,y_m)$.
Algebraically, 
\begin{equation}
\label{eq:R_m}
R_m(\mathbf y) \propto \star \left(\omega(y_1) \wedge \cdots \wedge \omega(y_m) \wedge \frac{\gamma^{\wedge (M-m)}}{(M-m)!}\right).
\end{equation}

This reduction is captured exactly by the \emph{Miwa time shift} \cite{Miwa1982}. 
We define the fundamental shift vector $[z^{-1}]$ by its components $[z^{-1}]_k = 1/(k z^k)$.
For a configuration $\mathbf{y}$, the total time deformation is the superposition 
$t_{\mathbf{y}} = -L^2 \sum_{i=1}^m [y_i^{-1}]$.
This time deformation corresponds to inserting $m$ charge-$L$ particles at
locations $\mathbf y$. The correlation function is therefore recovered by evaluating the $(M-m)$-particle 
$\tau$-function at these shifted times:
\[
  R_m(\mathbf{y}) = \prod_{i<j} |y_i - y_j|^{\beta} \prod_{i=1}^m e^{-V(y_i)} 
  \cdot \frac{\Tau_{M-m}(t_{\mathbf{y}})}{\Tau_M(0)}.
\]
Thus, particle coordinates act as discrete Miwa time parameters \cite{SinclairWells}. 
Inserting particles into the ensemble is equivalent to shifting the $\tau$-function 
along the momentum spine by the vector $t_{\mathbf{y}}$, bridging the gap between 
static probability and dynamic time evolution.

Equivalently, by the operator lifting machinery, the differential operator formed from the polynomial
$Q_{\mathbf y}(x) = \prod_i (x - y_i)^\beta$ corresponds to particle insertions, 
\[
  R_m(\mathbf y;t)
  = \prod_{i<j} |y_i - y_j|^{\beta} \prod_{i=1}^m e^{\xi(y_i, t)} 
  \cdot \frac{\widehat Q_{\mathbf y}(\partial)[\Tau_{M-m}(t)]}{\Tau_M(t)}.
\]
Both perspectives—Miwa specialization and differential operators—yield explicit 
formulas for all correlation functions in terms of the universal $\tau$-function. 

\textbf{Integrable structure.}
The Hirota bilinear identity asserts a bilinear relationship between particle
insertion and annihilation in a pair of systems of delocalized particles,
mediated by conjugate Miwa shifts. Consider the pair of backgrounds
$\gamma(t)^{\wedge(M-1)} \otimes \gamma(t')^{\wedge(M+1)}$ representing $M-1$
(respectively $M+1$) delocalized particles in potentials determined by times $t$
and $t'.$ At the algebraic level, particle insertion corresponds to wedging with
a fixed $L$-blade, while removal corresponds to contraction with respect to a
dual blade; the bilinear identities arise from the fact that these operations
must satisfy the momentum conserving Pl\"ucker relations in the momentum spine.
The Hirota identity here is not an added assumption but a bookkeeping identity
enforcing momentum conservation under paired particle insertion and removal.

Insertion of a particle into the background at time $t$ produces a distribution
over the momentum labels of the inserted particle. If the new particle occupies
momentum sector $j$, then momentum conservation restricts the background states
of $\gamma(t)^{\wedge(M-1)}$ to total momentum $-j$. The corresponding probability
generating function is
\[
e^{\xi(z,t)}\frac{\Tau_{M-1}(t-L^2[z^{-1}])}{\Tau_M(t)}
\;\propto\;
\star\bigl(\omega(z)\wedge\gamma(t)^{\wedge(M-1)}\bigr).
\]
Particle insertion is thus mediated by the negative Miwa shift $-L^2[z^{-1}]$.

Particle annihilation in the background $\gamma(t')^{\wedge(M+1)}$ is generated by
the conjugate positive Miwa shift. As shown in the Supplemental Material, there
exists a blade $\Omega(z,t')$ in the dual momentum spine such that the increment
$\Delta\Tau_{M+1}(z, t')=\Tau_{M+1}(t'+L^2[z^{-1}])-\Tau_{M+1}(t')$ satisfies
\[
\Delta\Tau_{M+1}(z, t') \propto \star \bigl( \iota_{\Omega(z,t')} \gamma(t')^{\wedge M}\bigr)
\]
Up to scalar normalization, this contraction generates the shift of
momentum–spine amplitudes induced by particle annihilation.

The Hirota bilinear identity asserts the vanishing of the constant coefficient
in the Laurent expansion,
\[
[z^0]\;
\Tau_{M-1}(t-L^2[z^{-1}])\,
\Delta\Tau_{M+1}(z, t'),
\]
which pairs momentum insertion and removal between the two systems, ensuring
overall momentum conservation. Algebraically, this corresponds to the operator
$\omega(z)$ acting by wedge insertion on $\gamma(t)^{\wedge(M-1)}$ paired with
contraction by $\iota_{\Omega(z,t')}$ on $\gamma(t')^{\wedge(M+1)}$.

The Pl\"ucker relations on the momentum spine are generated by the identity
$\omega(z)\wedge\omega(z)=0$. This lifts to the vanishing of the combined
insertion–contraction operator $\omega(z) \otimes \iota_{\Omega(z,t')}$ acting
on $\gamma(t)^{\wedge(M-1)} \otimes \gamma(t')^{\wedge(M+1)}$, yielding the
finite $M$ Hirota bilinear identity. When the momentum band is extended and the
fixed-particle-number projection is removed, this bilinear identity recovers the
usual Hirota equations of the KP hierarchy with Miwa times scaled by $L^2$.

Although the resulting identities parallel those of the KP hierarchy, they are
derived here in a finite-dimensional setting and reflect exact momentum
conservation rather than an appeal to infinite-rank integrable structure.

\begin{figure}[t]
  \centering
  \includegraphics[width=\columnwidth]{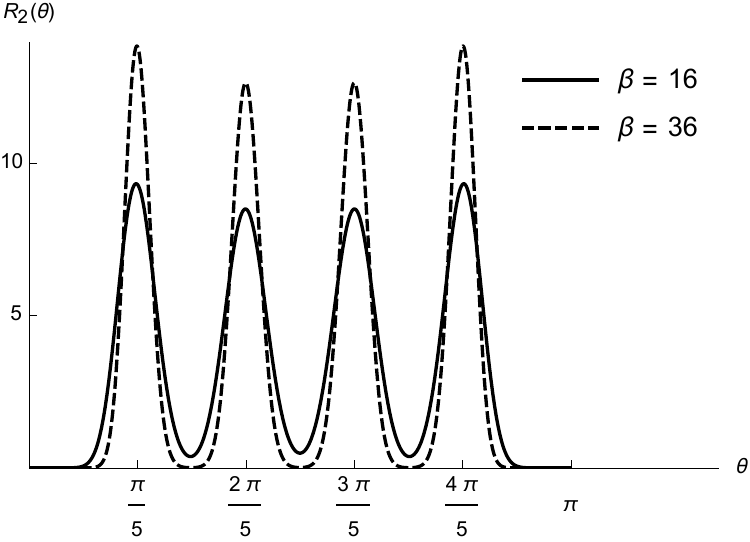}
  \caption{
    Exact two-point correlation function $R_2(\theta) := R_2(e^{i \theta/2}, e^{-i \theta/2})$ in the circular ensemble for $M=5$,
    shown for $\beta=16$ (solid) and $\beta=36$ (dashed).
    The normalization is fixed so that $\int_{0}^{\pi} R_2 \,d\theta=\binom{5}{2}=10$.
    As $\beta$ increases, the peaks sharpen at the nontrivial fifth roots of unity,
    consistent with increasing localization of relative particle separations.
  }
  \label{fig:R2_M5_beta16_beta36}
\end{figure}

\textbf{Circular Ensembles.}
The circular ensemble provides a canonical example in which the momentum
structure simplifies maximally. For Haar measure on the unit circle, the Gram
form collapses to a pure momentum state $\gamma(0) = \epsb_0.$ In contrast with
generic ensembles where all momentum sectors contribute, delocalized particles
in the circular ensemble populate only the momentum-zero sector. This makes
calculations particularly tractable, see Figure~\ref{fig:R2_M5_beta16_beta36} for exact
2-point functions for circular ensembles when $\beta=16, 36.$

This reveals an uncertainty principle. Complete delocalization on the circle
produces sharp momentum (a single sector $\epsb_0$), whereas inserting a
particle at $y$ activates all momentum sectors in the momentum spine. Sharp
momentum and sharp position are thus mutually exclusive extremes.

\textbf{Discussion.}
We have established that $\beta=L^{2}$ ensembles with $L$ even form an exactly
solvable class governed by the KP hierarchy. The integrable hierarchy arises at
finite $M$, without invoking asymptotic analysis. The partition function is a
$\tau$-function admitting a finite-dimensional Grassmannian interpretation,
satisfying the Hirota bilinear identity, with correlation functions generated by
partial differential operators acting on $\tau$.

This framework clarifies the relationship between classical and nonclassical
ensembles. For the classical symplectic case ($\beta=4, L=2$), the solvability
is guaranteed by the Darboux theorem, which allows the Gram 2-form to be (skew)
diagonalized globally. For $L > 2$, no such analogue of the Darboux theorem
exists for general forms, which has historically obscured the integrable
structure of higher-$\beta$ ensembles. The $\beta=L^2$ ensembles circumvent this
geometric obstruction because they do not involve generic $L$-forms, but rather
\emph{Wronskian} $L$-forms. The rigid algebraic structure of the Wronskian
defines a specific solvable substratum of the KP hierarchy corresponding to Miwa
times scaled by $L^2$.

Integrability is not imposed but emerges from the rigid combinatorics of
particle insertions constrained by momentum conservation. This emergent
conservation law acts as a selection rule, effectively reducing the
infinite-dimensional KP flows to the specific sectors compatible with the
$\beta=L^2$ interaction. This perspective suggests a meta-theory of ``charge
$L$'' hierarchies, where the classical symplectic ensembles ($L=2$) are the
basepoint of an infinite family of hyperpfaffian-solvable systems subsumed by a
Berezin-type calculus allowing for solvability of multi-component log-gases.
Future work will address the odd-$L$ case, which requires a related formalism of
$2L$-forms, and the large-$M$ asymptotics of these nonclassical integrable
systems.

\bibliographystyle{apsrev4-2}
\bibliography{references}

\onecolumngrid
\newpage
\begin{center}
\textbf{\large Supplemental Material for ``Exact Solvability via the KP Hierarchy for $\beta=L^{2}$ Random Matrix Ensembles''}
\end{center}

\setcounter{equation}{0}
\setcounter{figure}{0}
\setcounter{table}{0}
\setcounter{page}{1}
\makeatletter
\renewcommand{\theequation}{S\arabic{equation}}
\renewcommand{\thefigure}{S\arabic{figure}}
\renewcommand{\bibnumfmt}[1]{[S#1]}
\renewcommand{\citenumfont}[1]{S#1}

\section{I. Combinatorics of the Momentum Spine}

The central mechanism for exact solvability is the reduction from the full
exterior algebra $\Lambda^{L}V$ (dimension $\binom{LM}{L}$) to the
\emph{momentum spine}. For a configuration $u=(u_1,\ldots,u_L)$, the Wronskian
\[
\mathrm{Wr}(x^{u_1},\ldots,x^{u_L})
=
\Delta_u\,x^{u_1+\cdots+u_L-\frac12 L(L-1)},
\qquad
\Delta_u=\prod_{j<k}(u_k-u_j),
\]
is homogeneous in $x$ and naturally graded by its total power.

Defining the recentered momentum
\[
p_u := \sum_{i=1}^L u_i - \overline{\Sigma},
\qquad
\overline{\Sigma}=\tfrac{1}{2}L(LM-1),
\]
we introduce the momentum-sector vectors
\begin{equation}
\label{eq:spine_def}
\epsb_j := \sum_{p_u=j} \Delta_u\, e_u,
\end{equation}
whose linear span defines the momentum spine $\mathcal S$. For a system of $M$
particles of charge $L$, admissible momenta satisfy $|j| \le L^{2}(M-1)/{2},$
corresponding to allowable configurations of fermionic indices. Thus the
effective dimension scales as $O(L^{2}M)$, a dramatic reduction from
$\binom{LM}{L}$. 

\section{II. Universal Structure Coefficients}

Let $\star$ denote the Hodge projection onto the determinantal line in
$\Lambda^{LM}V$. For a momentum configuration
$\mathbf{j}=(j_1,\ldots,j_M)$ define $C_{\mathbf{j}} :=
\star\!\left(
\epsb_{j_1}\wedge\cdots\wedge\epsb_{j_M}
\right).$ Since the determinantal line carries zero net momentum,
\begin{equation}
C_{\mathbf{j}}=0
\qquad\text{unless}\qquad
\sum_{k=1}^M j_k = 0.
\end{equation}

Writing the time-dependent moments as
\[
m_j(t) := \int x^{P+j} e^{\xi(x,t)}\, d\mu(x),
\qquad
P=\tfrac{1}{2}L^{2}(M-1),
\]
the $\tau$-function admits the finite expansion
\begin{equation}
\tau_M(t)
=
\sum_{\mathbf{j}:\,\sum j_k=0}
C_{\mathbf{j}}
\prod_{k=1}^M m_{j_k}(t),
\end{equation}
separating universal interaction geometry from time dependent/ensemble-specific confinement.

\section{III. Pl\"ucker Relations on the Momentum Spine}

Let $V$ have dimension $L(M+1)$ and let $\mathcal S\subset\Lambda^{L}V$ denote
the momentum--spine subspace with basis $\{\epsb_j\}_{j=-P}^{P}$, where
$P=\tfrac12 L^2 M$.  Define the generating $L$--form
\begin{equation}\label{eq:omega}
  \omega(z):=\sum_{|j|\leq P} z^{P+j}\,\epsb_j .
\end{equation}
As a Wronskian $L$--blade, $\omega(z)$ is decomposable, hence
$\omega(z)\wedge\omega(z)=0$.  Extracting coefficients yields the quadratic
Pl\"ucker relations intrinsic to the momentum spine,
\begin{equation}\label{eq:plucker}
  \sum_{j+k=n} \epsb_j\wedge\epsb_k = 0 .
\end{equation}

Let $\langle \cdot , \cdot \rangle$ denote the canonical pairing between the
momentum spine $\mathcal S$ and its dual momentum spine $\mathcal S^*$. Under
this pairing, wedging by an element of $\mathcal S$ is adjoint to contraction by
the corresponding element of $\mathcal S^*$. Let
$\{\epsb_j^*\}\subset\mathcal S^*$ denote the fixed dual momentum--spine
basis, defined by $\epsb_j^*(\epsb_k)=\delta_{jk}$, and write
$\iota_{-j}:=\iota_{\epsb_{j}^*},$ so that $\iota_{-j}$ removes momentum $j$ from
the background by contraction. Then
\[
\langle \epsb_j\wedge A,\,B\rangle=\langle A,\,\iota_{-j} B\rangle.
\]
Pairing~\eqref{eq:plucker} with backgrounds
$\Psi\in \mathcal S^{\wedge(M-1)}$ and
$\Phi \in \mathcal S^{\wedge(M+1)}$ gives
\[
\sum_{k-j=n}\langle \epsb_k\wedge\Psi,\ \iota_j\Phi\rangle=0 ,
\]
which we refer to as the lifted Pl\"ucker relations.

\section{IV. Hirota Bilinear Identity}

The negative Miwa shift $t\mapsto t-L^2[z^{-1}]$ inserts a single particle and
produces
\[
\Tau_{M-1}(t-L^2[z^{-1}])
\propto
\star\!\bigl(\omega(z)\wedge\gamma(t)^{\wedge(M-1)}\bigr).
\]
Dually, the positive Miwa shift inserts a single hole.  At the Gram level this
acts by
\[
\gamma(t'+L^2[z^{-1}])=\gamma(t')+\Omega(z,t') ,
\]
where $\Omega(z,t')\in\mathcal S^*$ is the induced dual
momentum--spine element encoding the response of $\gamma(t')$ to the removal of a
particle at $z$.  Writing this element in the dual basis,
\begin{equation*}
\label{eq:Omega}
\Omega(z,t')=\sum_{k=1}^{2P} z^{-k}\,\varphi_k(t') ,
\qquad
\varphi_k(t')
=
\binom{L^2+k-1}{k}
\sum_j m_{j+k}(t')\,\epsb_j^* ,
\end{equation*}
reflecting the action of the positive Miwa shift on the moment sequence.

Define the Toeplitz linear operator
$\kappa(z,t') : V \rightarrow V^*$ with symbol $(1-x/z)^{-L^2}$.
Applying this operator componentwise to each factor of
$\omega(z)=v(z)\wedge v'(z)\wedge\cdots\wedge v^{(L-1)}(z)$, and expanding in the
dual basis shows
\[
\Omega(z,t') =
\kappa(z,t')v(z)\wedge \kappa(z,t')v'(z)\wedge\cdots\wedge
\kappa(z,t')v^{(L-1)}(z)
\;\in\; \Lambda^L V^* .
\]
Since $\Omega(z,t')$ is the image of a decomposable blade under a linear map, it
is itself decomposable, and therefore $\Omega(z,t')\wedge\Omega(z,t')=0$
identically. Since all higher--order insertions vanish, expanding $(\gamma(t') +
\Omega(z,t'))^{\wedge(M+1)}$ yields
\begin{equation}
\Delta\Tau_{M+1}(z, t')
\propto
\star\!\left(
\Omega(z,t')\wedge\frac{\gamma(t')^{\wedge M}}{M!}
\right).
\end{equation}

The scalar Laurent series
\begin{equation}
\mathcal H(z):=
\star \bigl(\omega(z)\wedge\gamma(t)^{\wedge(M-1)}\bigr)\,
\star\!\bigl(\Omega(z,t')\wedge\gamma(t')^{\wedge M}\bigr).
\end{equation}
represents the insertion of a single particle into the background $t$, paired
with the insertion of a single hole into the background $t';$ powers of $z$
generating the resulting total momentum change.

Expanding $\mathcal H(z)$ using the momentum--spine bases
$\{\epsb_k\}\subset\mathcal S$ and
$\{\varphi_j\}\subset\mathcal S^*$ gives
\[
\mathcal H(z)
=
\sum_{k,j} z^{k-j}
\,
\bigl\langle
\epsb_k\wedge\gamma(t)^{\wedge(M-1)},
\,
\iota_{\varphi_j}\gamma(t')^{\wedge(M+1)}
\bigr\rangle .
\]
Writing $\varphi_j$ in the dual basis, the lifted Pl\"ucker relations imply that, for each fixed
integer $n$, the sum of these pairings over $k-j=n$ vanishes.  In particular, the
coefficient of $z^0$ is $[z^0]\mathcal H(z)=0.$ This is equivalent to 
\begin{equation}
[z^0]\; \Tau_{M-1}(t-L^2[z^{-1}])\,\Delta\Tau_{M+1}(z, t')=0
\end{equation}
which is precisely the finite-$M$ Hirota bilinear identity for the
$\beta=L^2$ ensemble, up to conventional normalization factors.

\end{document}